\documentclass[preprint]{aastex}
\newcommand{\ha}{\mbox{H$\alpha$}}

\newcommand{\oiiir}{\mbox{[\ion{O}{3}]~$\lambda$5007}}

\singlespace
\slugcomment{Submitted to the Astronomical Journal}
\begin{document}

\title{Tip of the Red Giant Branch Distances to \objectname{NGC 4214},
\objectname{UGC 685}, and \objectname{UGC 5456}\altaffilmark{1}} 
\shorttitle{Distances to NGC 4214, UGC 685, and UGC 5456}

\author{Jes\'us Ma\'{\i}z-Apell\'aniz\altaffilmark{2},
        Lucas Cieza\altaffilmark{3}, and
        John W. MacKenty\altaffilmark{2}}


\altaffiltext{1}{Based on observations with the NASA/ESA {\em Hubble Space 
Telescope} obtained at the Space Telescope Science Institute, which is 
operated by the Association of Universities for Research in Astronomy, Inc. 
under NASA contract No. NAS5-26555.}

\altaffiltext{2}{Space Telescope Science Institute, 3700 San Martin Drive, 
Baltimore, MD 21218, U.S.A.}

\altaffiltext{3}{Department of Physics and Space Sciences, Florida Institute of
Technology, Melbourne, FL 32901-6975, U.S.A.}

\begin{abstract}
	We have used WFPC2 $VRI$ observations to calculate the distances to
three nearby galaxies, NGC 4214, UGC 685, and UGC 5456 using the tip of the red
giant branch method. Our values for NGC 4214 ($2.94\pm 0.18$ Mpc) and UGC 685
($4.79\pm 0.30$ Mpc) are the most precise measurementes of the distances to 
these objects ever made. For UGC 5456 the data do not allow us to reach a 
decisive conclusion since there are two possible solutions, one leading towards 
a short distance around 3.8 Mpc and another one towards a long distance of 5.6 
Mpc or more.
\end{abstract}

\keywords{galaxies: distances and redshift --- 
galaxies: individual (NGC 4214, UGC 685, UGC 5456) --- galaxies: irregular ---
galaxies: stellar content}

\section{INTRODUCTION}

	In the last decade, the measurement of the tip of the red giant branch 
(or TRGB) has become a reliable method for measuring distances to galaxies with 
resolved stellar populations \citep{Leeetal93}. This has been posible thanks to
the very weak dependence of the absolute magnitude of the TRGB on metallicity
\citep{DaCoArma90,Belletal01}. Another advantage of this method (as opposed to
the use of variable stars such as Cepheids or RR Lyrae) is that only a single 
epoch is needed to estimate the distance, considerably reducing the amount of
observing time required and avoiding scheduling problems. Its robustness
has been recently tested by obtaining consistent distances to the SMC, 
LMC, and IC 1613 using four different methods: Cepheids, RR Lyrae, red clump, 
and TRGB \citep{Dolpetal01a}. The use of HST/WFPC2 data allows the
measurement of distances up to $\approx 5$ Mpc with a single orbit using the 
TRGB \citep{Karaetal01,Tosietal01,Dolpetal01b} and the introduction of ACS will 
extend that range in distance by a factor of two. 

	The method for using the TRGB as a standard candle has been described by
\citet{Sakaetal96}. One constructs a smoothed version of the luminosity 
function in the Cousins $I$ passband as:

\begin{equation}
\Phi(I) = \sum_{i=1}^{N} \frac{1}{\sqrt{2\pi}\sigma_i}
\exp\left[-\frac{(I_i-I)^2}{2\sigma_i^2}\right], \label{Phi}
\end{equation}

\noindent where $I_i$ and $\sigma_i$ are the magnitude and photometric
uncertainty of the $i$th star, respectively, and $N$ is the total number of
stars in the sample. We then define an adaptive edge-detection filter as:

\begin{equation}
E(I) = \Phi(I+\bar{\sigma}_m) - \Phi(I-\bar{\sigma}_m), \label{E}
\end{equation}

\noindent where $\bar{\sigma}_m$ is the mean photometric uncertainty for all
stars with magnitudes between $I-0.05$ and $I+0.05$. $E$ can then be used as a
localized slope estimator to find the cutoff in the $I$ band luminosity function
created by the TRGB.

	The three objects in this paper (NGC 4214, UGC 685, and UGC 5456) were 
selected from a study of a complete sample of galaxies from the Center for 
Astrophysics redshift survey (CfaRS; \citealt{Huchetal83b}) presented by 
\citet{Burg87} \footnote{The CfaRS consisted of 2\,400 galaxies taken from the 
original Zwicky Catalog to a limiting magnitude of $M_B^0 = 14.5$.} on the 
basis of their high excitation and proximity. A summary of the properties of the
three galaxies is given in Table~\ref{reference}. 

	NGC 4214 is a well-studied nearby magellanic irregular galaxy. It has
several intense star-forming regions concentrated along the bar of the galaxy
\citep{MacKetal00,Maiz01b} which can be easily studied thanks to the low
foreground extinction and the thin character of its galactic disk
\citep{Maizetal99a}. Furthermore, the different massive young clusters are at
different evolutionary stages \citep{Maizetal98}. These characteristics make NGC
4214 the best available template for dwarf starbursts. \citet{Leitetal96}
estimate its distance as 4.1 Mpc but they admit that the value could be lower if
the distance to the Virgo Cluster is different than the one they use for their
calculations. They also cite other sources' estimates ranging as high as 
5.1 Mpc. On the other hand, \citet{Hoppetal99} measure a distance of 
$\approx$ 2 Mpc using NICMOS data to find the TRGB.

	UGC 685 and UGC 5456 are two smaller and less studied galaxies. They are
both classified as Blue Compact Dwarfs (or BCDs) and show regions of current
star formation activity, though not as intense as those in NGC 4214. 
\citet{Hopp99} and \citet{MakaKara98} find values of 5.5 Mpc and 5.9 Mpc,
respectively, for the distance to UGC 685 based on the brightest blue 
supergiant stars. \citet{MakaKara98} also find a distance to UGC 5456 of 2.7 Mpc
using the same method but specify that this value is in apparent disagreement 
with its radial velocity.

\section{OBSERVATIONS AND DATA REDUCTION} 

 	We obtained deep, high resolution, multiwavelength 
imaging of NGC 4214, UGC 685, and UGC 5456 with the WFPC2 instrument aboard HST 
(prop. ID 6569) on 1997 July 22, 1998 Nov 17, and 1999 Feb 18, respectively.
Four continuum filters (F336W, WFPC2 $U$; F555W, WFPC2 $V$; F702W, WFPC2 wide
$R$; and F814W, WFPC2 $I$) and two nebular filters (F656N, \ha\ and F502N,
\oiiir) were used for each galaxy, as shown in Table~\ref{wfpc2obs}. The 
NGC 4214 nebular data and the structure of its stellar clusters were analyzed 
in two previous papers \citep{MacKetal00,Maiz01b}; the detailed study of the
young population will be presented in a follow-up paper \citep{Maizetal02a}.
In this paper we analyze the F555W, F702W, and 
F814W data of the three galaxies in order to measure their distances by finding 
the tip of the red giant branch.

	The reduction of the NGC 4214 data was described in \citet{MacKetal00} 
and the reader is referred to that paper for details. The UGC 685 and UGC 5456 
data were reduced in an analogous way. In particular, we point out that the 
F502N and F656N images were used to eliminate the nebular contribution to the 
F555W and F702W data, thus producing ``near-pure'' $V$ and $R$ continuum images.
With one exception (the F702W filter for NGC 4214), two long and one short
exposures were obtained for each filter and galaxy in order to eliminate cosmic 
rays and to avoid saturation at the center of compact bright sources. 

	We obtained the 4-band stellar photometry of each galaxy using the 
HSTphot package \citep{Dolp00a}. HSTphot is tailored to handle the undersampled
nature of the PSF in WFPC2 images and uses a self-consistent treatment of the
CTE and zero-point photometric calibrations \citep{Dolp00b}. The central
regions of the three galaxies are dominated by a high surface density young
stellar population. In those areas, blending between stars is a serious 
problem due to crowding and isolating the old population is not possible even at
WFPC2 resolution. Therefore, we masked out those regions as well as the 
prominent NGC 4214 clusters identified by \citet{Maiz01b}. For the remaining 
areas, we eliminated those objects with reduced $\chi^2 > 4.0$ (poor fit:
likely multiple or extended objects) and sharpness\footnote{HSTphot defines
sharpness in such a way that ``perfect'' stars have a value of 0, more
centrally-concentrated objects (e.g. cosmic-rays) have a negative value and
less-concentrated ones (e.g. galaxies) have a positive value.} $< -0.3$ or
$> 0.3$ and the resulting data was grouped in three sets for each galaxy: [1]
stars detected in all four filters (1\,595, 325, and 197 stars for NGC 4214, UGC
685, and UGC 5456, respectively), [2] stars detected in at least F555W and F814W
(13\,408, 3\,594, and 1\,105 stars, respectively), and [3] stars detected in at 
least F702W and F814W (17\,936, 5\,207, and 1\,803 stars, respectively). For
each galaxy, the data in the first set were used to measure extinction (as
described later) while the data in the second and third sets were used to 
detect the tip of the red giant branch in each galaxy.

\begin{figure}
\centerline{\includegraphics*[width=\linewidth]{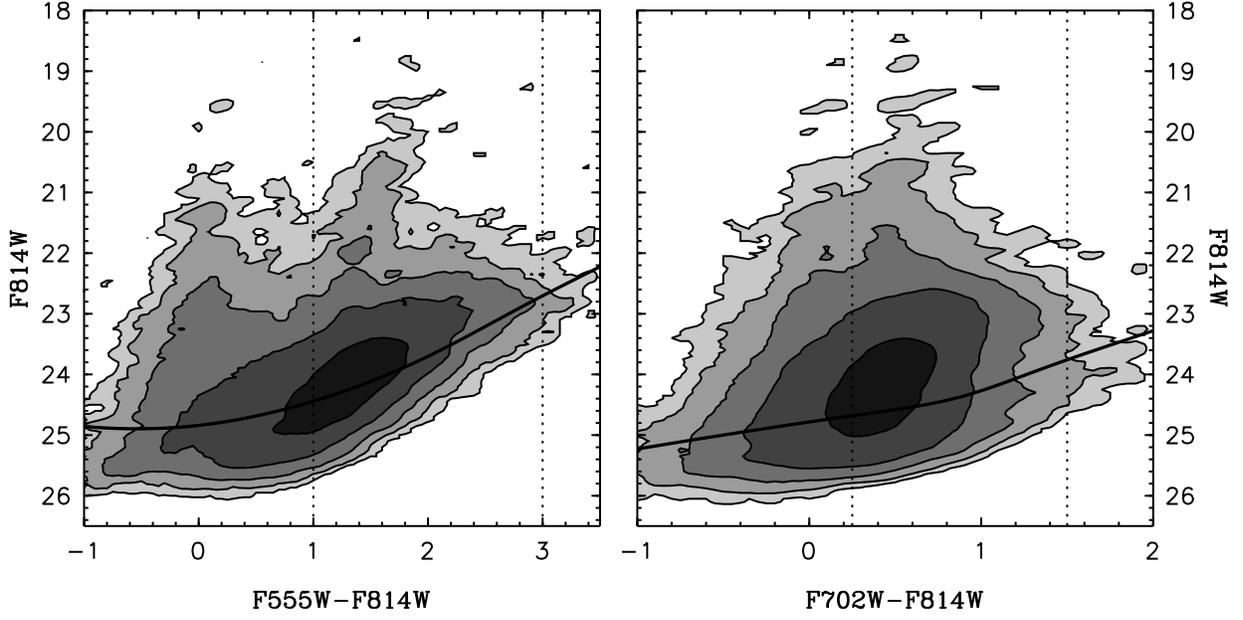}}
\caption{Observed color-magnitude contour plots of NGC 4214. Stars in the 
central regions and in prominent young clusters are excluded in order to reduce 
blending and enhance the old population. Contours are logarithmically spaced.
The thick solid line marks the 50\% completeness limit but the plots themselves
are not corrected for that effect. The vertical dashed lines indicate the color
range used to determine the location of the tip of the red giant branch.}
\label{cmdsngc4214}
\end{figure}

\begin{figure}
\centerline{\includegraphics*[width=\linewidth]{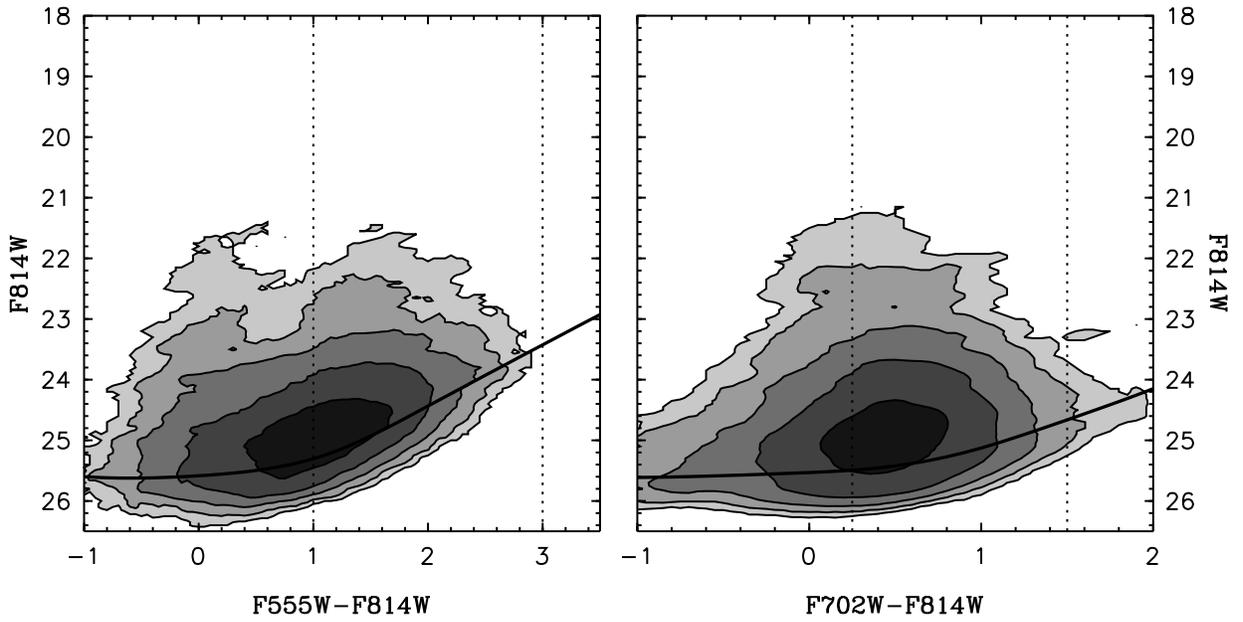}}
\caption{Same as Fig.~\ref{cmdsngc4214} for UGC 685.}
\label{cmdsugc685}
\end{figure}

\begin{figure}
\centerline{\includegraphics*[width=\linewidth]{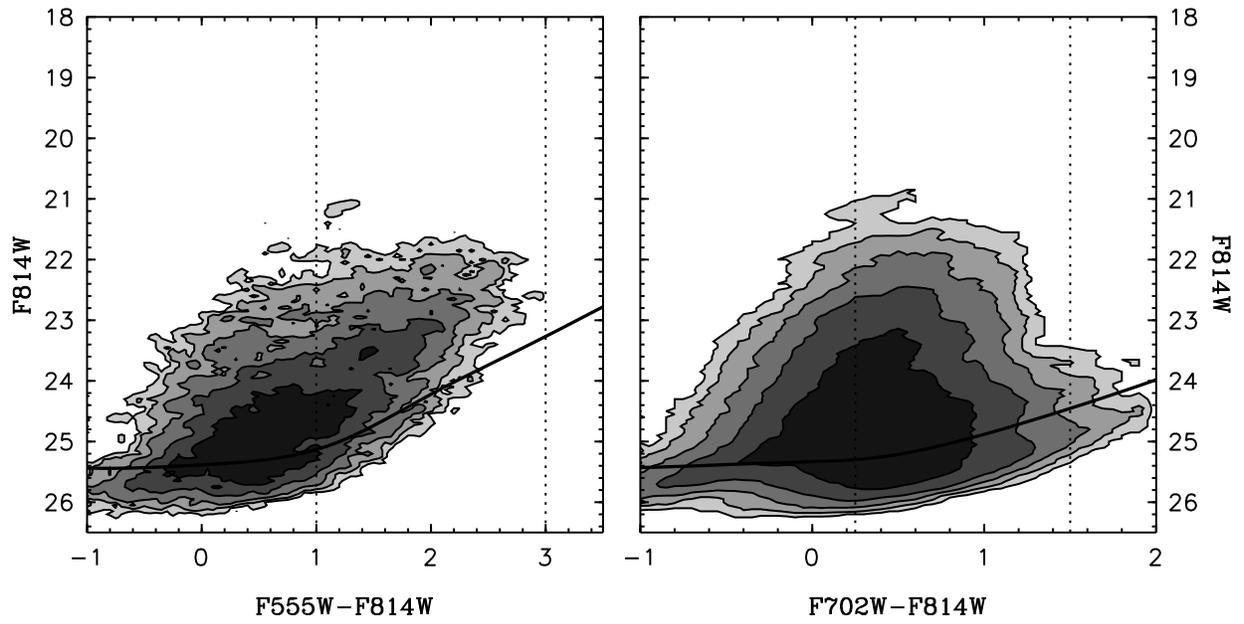}}
\caption{Same as Fig.~\ref{cmdsngc4214} for UGC 5456.}
\label{cmdsugc5456}
\end{figure}

	The F555W$-$F814W vs. F814W and F702W$-$F814W vs. F814W color-magnitude
contour plots are shown in 
Figs.~\ref{cmdsngc4214},~\ref{cmdsugc685},~and~\ref{cmdsugc5456}. Contours
are logaritmically spaced (in stars / color magnitude / brightness 
magnitude) in order to show the structures at both low and high
densities. For each of the six possible combinations (three galaxies and two 
color-magnitude pairs) we performed artificial star experiments using the 
{\it hstfake} utility available in HSTphot. In each case experiments were 
performed in six 0.25 color-magnitudes-wide ranges and the results were 
interpolated for intermediate colors. In 
Figs.~\ref{cmdsngc4214},~\ref{cmdsugc685},~and~\ref{cmdsugc5456} we show the
location of the 50\% completeness limit as a function of color.

\section{ANALYSIS}

\begin{figure}
\centerline{\includegraphics*[width=\linewidth]{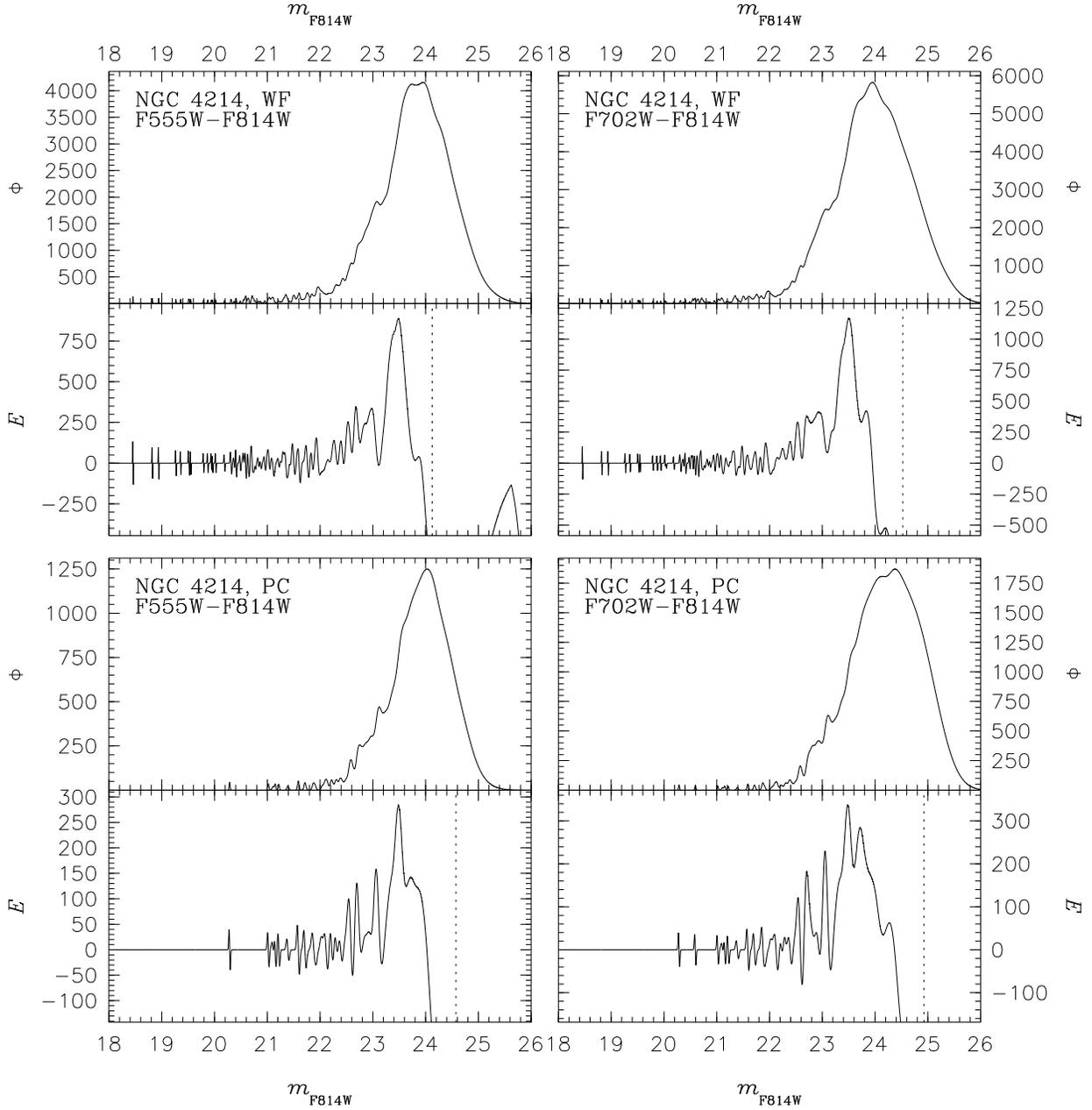}}
\caption{F814W luminosity functions with the corresponding outputs of the
edge-detection filter for NGC 4214. The upper plots are for the non-masked
regions in the WF chips while the lower ones are for the PC chip. The left plots
use the F555W data for the color information while the right ones use the F702W
data. The dashed vertical line marks the 50\% completeness in each case.}
\label{finalngc4214}
\end{figure}

\begin{figure}
\centerline{\includegraphics*[width=\linewidth]{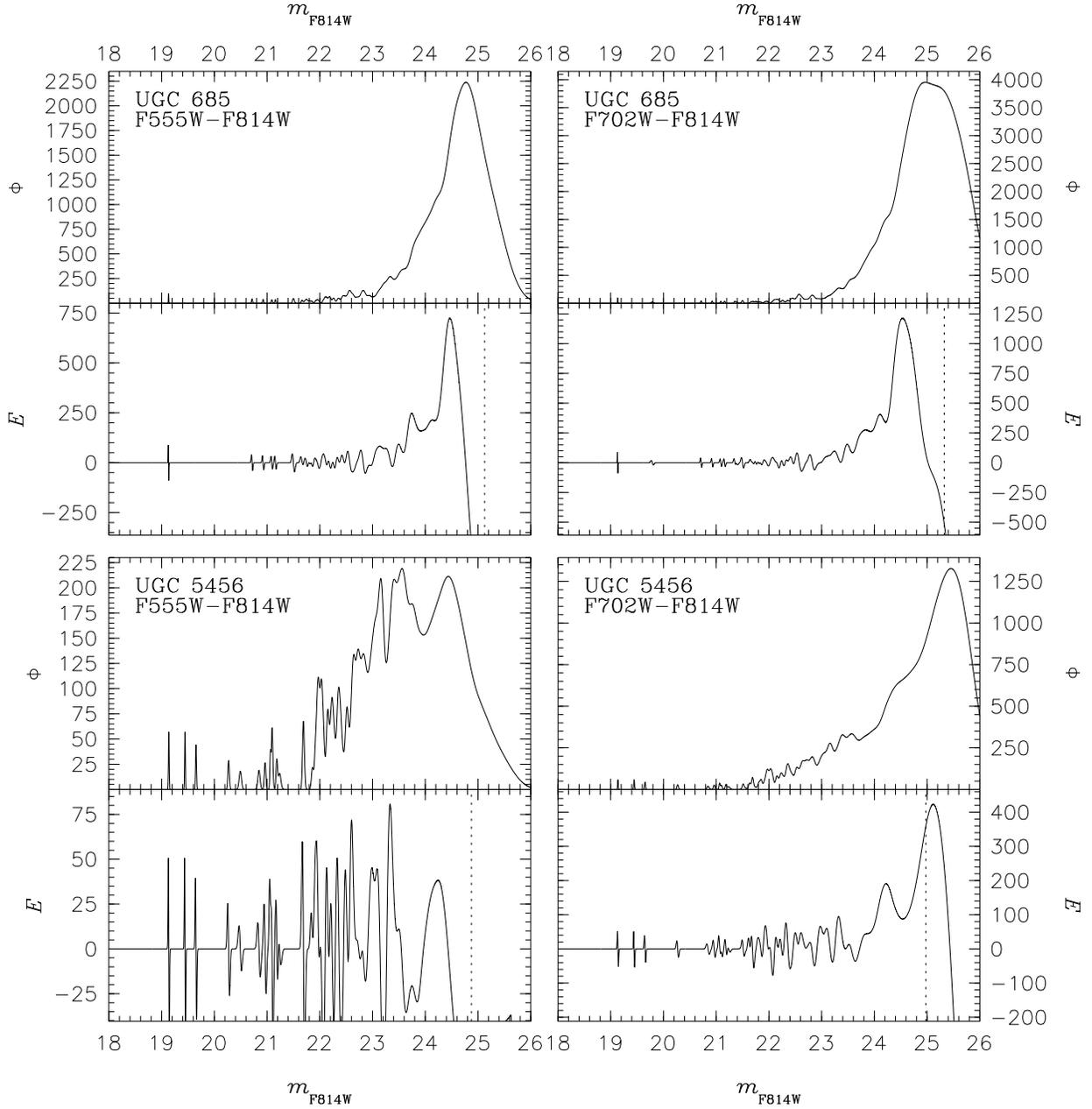}}
\caption{F814W luminosity functions with the corresponding outputs of the
edge-detection filter for UGC 685 (upper plots) and UGC 5456 (lower plots). The 
left plots use the F555W data for the color information while the right ones 
use the F702W data. The dashed vertical line marks the 50\% completeness in 
each case.}
\label{finalugc685ugc5456}
\end{figure}

	We selected those stars with F555W$-$F814W between 1.0 and 3.0 in the
second set and those with F702W$-$F814W between 0.25 and 1.5 in the third set 
and we built the corresponding smoothed luminosity functions in the F814W
passband according to Equation~\ref{Phi}. The values were chosen in order to 
maximize the number of red giants included while minimizing the number of stars
of other types using \citet{DaCoArma90} as a reference for $V-I$ colors and
the \citet{Lejeetal97} models to transform them to $R-I$. 
We then calculated the corresponding edge-detection filters
according to Equation~\ref{E}. The results are shown in 
Figs.~\ref{finalngc4214}~and~\ref{finalugc685ugc5456} and the location of the
TRGB in each case is shown in Table~\ref{distances}. The values of $\Phi$ and 
$E$ shown in Figs.~\ref{finalngc4214}~and~\ref{finalugc685ugc5456} have been 
corrected for incompleteness. However, it should be noted that the
incompleteness correction can become very uncertain for faint magnitudes due to
small-number statistics. For that reason, we have marked in each case the 
location of the weighted 50\% incompleteness limit. Only the data to the left 
of that point should be trusted when finding the TRGB.

	In order to test the effects of crowding we separated the NGC 4214 data
in two subsets for each of the two band sets depending on the chip where the
stars were detected (PC or WF2+3+4). Since the PC has a pixel size smaller than
the WF chips by a factor of 2.2, it provides a considerably better sampling of
the PSF, so it should alleviate any possible crowding problems. However, we do
not find any significant difference between the PC and the WF results. In all
four cases a clear peak appears at essentially the same value of $m_{\rm F814W}$
and only for the F702W$-$F814W PC subset we do find a 0.3 magnitudes fainter 
secondary peak. Such a secondary peak would lead to an undecisive conclusion
regarding the position of the TRGB if that were our only data. However, the
availability of the results from the other three subsets allows us to establish
the magnitude of the TRGB with precision.

	For UGC 685 and UGC 5456 there are not enough stars in the PC chip to 
perform a similar experiment. Nevertheless, the UGC 685 data
allow us to see the differences between using both sets. There are
approximately twice as many stars detected in F702W and F814W than in F555W and
F814W due to two reasons: red stars are more easily detected in F702W than in
F555W and the lower separation in color introduces more ``blue'' stars in the
luminosity function. The first reason is an argument in favor of using F702W
instead of F555W (more real red giants are detected) while the second one is an
argument against it (more spureous objects are included in the sample), so the
simultaneous use of the two sets (F555W$-$F814W and F702W$-$F814W) is a good
check of the validity of the results. For UGC 685, the edge-detection filters
for the two sets yield a consistent value for the TRGB which is well to the 
left of the 50\% completeness limit, so a distance measurement can be decisively
established.

	Unfortunately, the same is not true for UGC 5456. A strong peak is
visible in the F702W$-$F814W edge detection filter at $m_{\rm F814W}=25.07$ 
but it is located to the right of the 50\% incompleteness limit and, therefore, 
cannot be trusted to be real (the peak is much weaker if the incompleteness 
correction is not applied to $\Phi$). A secondary peak is visible at 
$m_{\rm F814W}=24.24$ in the same data set and also at $m_{\rm F814W}=24.20$ in
the F555W$-$F814W data set which could be caused by the TRGB but we cannot be
certain about it\footnote{The F555W$-$F814W UGC 5456 
data set is a good example of one of
the problems in the TRGB method: One needs to detect a large number of bright
red giants in order to minimize Poisson fluctuations in $\Phi$ near the location
of the TRGB. This is especially problematic when a sizable population of red
super giants is present.}. Therefore, we conclude that UGC 5456 either is 
farther away than the range that can be measured with the present data or has 
very few red giants outside its central regions.

	In order to arrive at a distance from our data there are three required
steps. First, we have to correct for extinction. A minimum value for 
$E(B-V)$ is provided by the foreground (Galactic) value measured for the 
position of each galaxy by \citet{Schletal98} from COBE and IRAS data (see
Table~\ref{extinctions}). However, the \citet{Schletal98} values do not take
into account internal extinction within the galaxy itself. In order to measure
that, we used the set with 4-band photometry and generated the three colors
F336W$-$F555W, F555W$-$F702W, and F555W$-$F814W, We then applied an extension 
to $>2$ colors of the standard color-color procedure (described in 
\citealt{Maizetal02a}) to calculate the median $E(B-V)$ which affects the
blue stellar population in each of the three galaxies outside the masked
regions (i.e. the blue population which is approximately co-spatial with the red
giants). The results are also shown in Table~\ref{extinctions} and they are
slightly higher than the values provided by \citet{Schletal98}, as expected in
the case of moderate internal extinction. Nevertheless, it is possible
that some of the extinction associated with the blue population is localized
around the individual stars themselves, so we will consider this second value
to be an upper bound to the extinction affecting the red giant population.
Therefore, we will use as our value for the extinction in each case the mean of 
the two results (foreground and blue population) and we will estimate the
uncertainty as half the difference between them. The last column in 
Table~\ref{extinctions} shows the adopted correction in the $I$ band using a
value of $R_V$ of 3.1. Under the ``other'' column in Table~\ref{extinctions} we 
also list the values of $E(B-V)$ measured from Balmer ratios in H\,{\sc ii} 
regions in the central regions of NGC 4214 and UGC 685. As expected, those
values are higher than the ones we use due to the considerable internal
extinction present in those regions. They are shown here to serve as caution
against using values obtained in such a way in similar studies of other 
galaxies.

	The second necessary step to arrive to a distance is the conversion from
WFPC2 F814W magnitudes to Cousins $I$ ones. As described by \citet{Holtetal95b},
such conversions are non-trivial for the cases of high extinction and/or certain
WFPC2 filters. Fortunately, that is not the case for the galaxies analyzed here
and the F814W filter. In this paper we follow \citet{Holtetal95b} and use a
correction of $-0.03\pm 0.01$ magnitudes. 

	Finally, we have to choose a value for $M_{I,{\rm TGRB}}$. As shown by
\citet{Belletal01}, if the [Fe/H] of the studied population is known in detail,
it is possible to calculate $M_{I,{\rm TGRB}}$ within a few hundredths of a
magnitude. We do not have such a knowledge for the red giants in our galaxies
but we do not really need that level of precision either, since the 
uncertainties from our procedure for calculating $m_{I,{\rm TGRB}}$ are already
of the order of 0.1 magnitudes. Thus, we can settle for a value
$M_{I,{\rm TGRB}} = 4.00 \pm 0.10$, which is valid for the range of [Fe/H]
between $-2.8$ and $-0.6$ \citep{Belletal01} and almost certainly includes the 
metallicities of the red giant population in our three galaxies. 

	The values for the distances are shown in Table~\ref{distances}. For NGC
4214, three of the four values of $m-M$ are identical so we used one of those 
for our adopted distance as well as the smallest of the three uncertainties (the
values are not completely independent from the statistical point of view so 
there is no easy way to reduce the estimated uncertainty by combining them). 
For UGC 685 we used
the values obtained with the F702W$-$F814W data simply because of the larger
number of stars in the sample but, alternatively, the F555W$-$F814W result of
$4.61\pm 0.30$ Mpc could also be used (the distance between them is of only
0.6$\sigma$). Finally, as previously mentioned, no decisive result could be
obtained for UGC 5456 so two possible values are listed. 

	Our distance for NGC 4214 ($2.94\pm 0.18$ Mpc) is in the middle of the 
range previously obtained by other authors (2.0 to 5.1 Mpc) but no
compatibility analysis can be provided since the sources do not provide
uncertainty estimates. Our distance for UGC 685 ($4.79\pm 0.30$ Mpc) is also
quite close to the previous values available from the literature (5.5 Mpc,
\citealt{Hopp99}, and 5.9 Mpc, \citealt{MakaKara98}). Our result is well within 
the uncertainty specified by the first of those articles, which is 30\% (the 
second paper does not provide an uncertainty estimate). On the other hand, the 
previously value available for UGC 5456 (2.7
Mpc, \citealt{MakaKara98}) can certainly be excluded with our data; that galaxy
is farther away, though we are not sure by how much. It is interesting to note
that the previous values for the distances to UGC 685 and UGC 5456 were obtained
using the same method, the magnitude of the brightest blue supergiant stars. Why
would the method be valid for the first galaxy but not for the second? The
explanation we prefer is that the brightest young clusters in UGC 685 are Scaled
OB Associations while those in UGC 5456 are Super Star Clusters (or at least
compact clusters), as described by \citet{Maiz01b}. Therefore, the brightest
point-like sources detected from the ground in $B$ in UGC 685 are likely to be 
individual blue supergiants or small aggregrates of them. On the other hand, the
brightest point-like sources in UGC 5456 are clusters made out of many young
stars, and confusing them with blue supergiants leads to a large underestimation
of the distance.

\acknowledgments

Lucas Cieza acknowledges support from the Space Telescope Science Institute
Summer Student Program. Support for this work was provided by NASA through 
grants GO-06569.01-A and GO-09096.01-A from the Space Telescope Science 
Institute, Inc., under NASA contract NAS5-26555. This research has made use of 
the NASA/IPAC Extragalactic Database (NED) which is operated by the Jet 
Propulsion Laboratory, California Institute of Technology, under contract with 
the National Aeronautics and Space Administration.

\bibliographystyle{aj}
\bibliography{general}

\begin{thebibliography}{}

\bibitem[\protect\citeauthoryear{Bellazini, Ferraro, \& Pancino}{Bellazini
  et~al.}{2001}]{Belletal01}
Bellazini, M., Ferraro, F.~R.,  \& Pancino, E. 2001, ApJ, 556, 635

\bibitem[\protect\citeauthoryear{Burg}{Burg}{1987}]{Burg87}
Burg, R.~I. 1987, Ph.D. thesis, Massachusetts Institute of Technology

\bibitem[\protect\citeauthoryear{Da~Costa \& Armandroff}{Da~Costa \&
  Armandroff}{1990}]{DaCoArma90}
Da~Costa, G.~S.,  \& Armandroff, T.~E. 1990, AJ, 100, 162

\bibitem[\protect\citeauthoryear{Dolphin}{Dolphin}{2000a}]{Dolp00a}
Dolphin, A.~E. 2000a, PASP, 112, 1383

\bibitem[\protect\citeauthoryear{Dolphin}{Dolphin}{2000b}]{Dolp00b}
Dolphin, A.~E. 2000b, PASP, 112, 1397

\bibitem[\protect\citeauthoryear{Dolphin et~al.}{Dolphin
  et~al.}{2001a}]{Dolpetal01b}
Dolphin, A.~E., et~al. 2001a, MNRAS, 324, 249

\bibitem[\protect\citeauthoryear{Dolphin et~al.}{Dolphin
  et~al.}{2001b}]{Dolpetal01a}
Dolphin, A.~E., et~al. 2001b, ApJ, 550, 554

\bibitem[\protect\citeauthoryear{Holtzman et~al.}{Holtzman
  et~al.}{1995}]{Holtetal95b}
Holtzman, J., Burrows, C.~J., Casertano, S., Hester, J.~J., Trauger, J.~T.,
  Watson, A.~M.,  \& Worthey, G. 1995, PASP, 107, 1065

\bibitem[\protect\citeauthoryear{Hopp}{Hopp}{1999}]{Hopp99}
Hopp, U. 1999, A\&AS, 134, 317

\bibitem[\protect\citeauthoryear{Hopp et~al.}{Hopp et~al.}{1999}]{Hoppetal99}
Hopp, U., Schulte-Ladbeck, R.~E., Greggio, L.,  \& Crone, M.~M. 1999, in ASP
  Conf. Series Proc., Vol. 192, 85

\bibitem[\protect\citeauthoryear{Huchra et~al.}{Huchra
  et~al.}{1983}]{Huchetal83b}
Huchra, J.~P., Davis, M., Latham, D.,  \& Tonry, J. 1983, ApJS, 52, 89

\bibitem[\protect\citeauthoryear{Jansen et~al.}{Jansen
  et~al.}{2000}]{Jansetal00}
Jansen, R.~A., Fabricant, D., Franx, M.,  \& Caldwell, N. 2000, ApJS, 126, 331

\bibitem[\protect\citeauthoryear{Karachentsev \& Makarov}{Karachentsev \&
  Makarov}{1996}]{KaraMaka96}
Karachentsev, I.~D.,  \& Makarov, D.~A. 1996, AJ, 111, 794

\bibitem[\protect\citeauthoryear{Karachentsev et~al.}{Karachentsev
  et~al.}{2001}]{Karaetal01}
Karachentsev, I.~D., et~al. 2001, A\&A, 375, 359

\bibitem[\protect\citeauthoryear{Lee, Freedman, \& Madore}{Lee
  et~al.}{1993}]{Leeetal93}
Lee, M.~G., Freedman, W.~L.,  \& Madore, B.~F. 1993, ApJ, 417, 553

\bibitem[\protect\citeauthoryear{Leitherer et~al.}{Leitherer
  et~al.}{1996}]{Leitetal96}
Leitherer, C., Vacca, W., Conti, P.~S., Filippenko, A.~V., Robert, C.,  \&
  Sargent, W. L.~W. 1996, ApJ, 465, 717

\bibitem[\protect\citeauthoryear{Lejeune, Buser, \& Cuisinier}{Lejeune
  et~al.}{1997}]{Lejeetal97}
Lejeune, T., Buser, R.,  \& Cuisinier, F. 1997, A\&AS, 125, 229

\bibitem[\protect\citeauthoryear{MacKenty et~al.}{MacKenty
  et~al.}{2000}]{MacKetal00}
MacKenty, J.~W., Ma\'{\i}z-Apell\'aniz, J., Pickens, C.~E., Norman, C.~A.,  \&
  Walborn, N.~R. 2000, AJ, 120, 3007

\bibitem[\protect\citeauthoryear{Ma\'{\i}z-Apell\'aniz}{Ma\'{\i}z-Apell\'aniz}%
{2001}]{Maiz01b}
Ma\'{\i}z-Apell\'aniz, J. 2001, ApJ, in press, available from {\tt
  http://www.stsci.edu/\~{}jmaiz}

\bibitem[\protect\citeauthoryear{Ma\'{\i}z-Apell\'aniz, Cieza, \&
  MacKenty}{Ma\'{\i}z-Apell\'aniz et~al.}{2002}]{Maizetal02a}
Ma\'{\i}z-Apell\'aniz, J., Cieza, L.,  \& MacKenty, J.~W. 2002, AJ, submitted

\bibitem[\protect\citeauthoryear{Ma\'{\i}z-Apell\'aniz
  et~al.}{Ma\'{\i}z-Apell\'aniz et~al.}{1998}]{Maizetal98}
Ma\'{\i}z-Apell\'aniz, J., Mas-Hesse, J.~M., {Mu\~noz-Tu\~n\'on}, C.,
  V\'{\i}lchez, J.~M.,  \& Casta\~neda, H.~O. 1998, A\&A, 329, 409

\bibitem[\protect\citeauthoryear{Ma\'{\i}z-Apell\'aniz
  et~al.}{Ma\'{\i}z-Apell\'aniz et~al.}{1999}]{Maizetal99a}
Ma\'{\i}z-Apell\'aniz, J., {Mu\~noz-Tu\~n\'on}, C., Tenorio-Tagle, G.,  \&
  Mas-Hesse, J.~M. 1999, A\&A, 343, 64

\bibitem[\protect\citeauthoryear{Makarova \& Karachentsev}{Makarova \&
  Karachentsev}{1998}]{MakaKara98}
Makarova, I.~N.,  \& Karachentsev, I.~D. 1998, A\&AS, 133, 181

\bibitem[\protect\citeauthoryear{Sakai, Madore, \& Freedman}{Sakai
  et~al.}{1996}]{Sakaetal96}
Sakai, S., Madore, B.~F.,  \& Freedman, W.~L. 1996, ApJ, 461, 713

\bibitem[\protect\citeauthoryear{Schlegel, Finkbeiner, \& Davis}{Schlegel
  et~al.}{1998}]{Schletal98}
Schlegel, D.~J., Finkbeiner, D.~P.,  \& Davis, M. 1998, ApJ, 500, 525

\bibitem[\protect\citeauthoryear{Tosi et~al.}{Tosi et~al.}{2001}]{Tosietal01}
Tosi, M., Sabbi, E., Bellazini, M., Aloisi, A., Greggio, L., Leitherer, C.,  \&
  Montegriffo, P. 2001, AJ, 122, 1271

\end{thebibliography}

\singlespace


\begin{deluxetable}{lccc}
\tablecaption{Reference data for the three galaxies. All data was obtained from
NED with the only exception of $v_{\rm LG}$, the velocity with respect to the
Local Group of galaxies, which was calculated from the heliocentric velocity, 
$v_{\rm helio}$ (obtained from NED), and the information provided by 
\citet{KaraMaka96}. Coordinates correspond to the J2000.0 epoch. 
\label{reference}}
\tabletypesize{\small}
\tablewidth{0pt}
\tablehead{ & \colhead{NGC 4214} & \colhead{UGC 685} &\colhead{UGC 5456}}
\startdata
RA                & $12^{\mathrm h}15^{\mathrm m}38\fs9$ & $01^{\mathrm h}07^{\mathrm m}22\fs4$ & $10^{\mathrm h}07^{\mathrm m}19\fs6$ \\
dec               & $+36\arcdeg19\arcmin40\arcsec$       & $+16\arcdeg41\arcmin02\arcsec$       & $+10\arcdeg21\arcmin46\arcsec$       \\
$v_{\rm helio}$   & 291 km s$^{-1}$                      & 157 km s$^{-1}$                      & 544 km s$^{-1}$                      \\
$v_{\rm LG}$      & 295 km s$^{-1}$                      & 351 km s$^{-1}$                      & 377 km s$^{-1}$                      \\
diameters         & $8\farcm5\times 6\farcm6$            & $1\farcm2\times 0\farcm9$            & $1\farcm6\times 0\farcm8$            \\
$m_{B}$           & $10.20\pm 0.16$                      & $14.23\pm 0.18$                      & $13.52\pm 0.18$                      \\
\enddata
\end{deluxetable}

\begin{deluxetable}{lllll}
\tablecaption{Observations.\label{wfpc2obs}}
\tabletypesize{\small}
\tablewidth{0pt}
\tablehead{\colhead{Filter} & \colhead{Band} & \colhead{Galaxy} & Images & 
Exp. time (s)}
\startdata
F656N & \ha            & NGC 4214 & u3n8010fm + gm      & 800 +    800          \\
      &                & UGC 685  & u3n8030fr + gr      & 800 +    800          \\
      &                & UGC 5456 & u3n8020fr + gr      & 400 +    400          \\
F502N & \oiiir         & NGC 4214 & u3n8010dm + em      & 700 +    800          \\
      &                & UGC 685  & u3n8030dr + er      & 700 +    800          \\
      &                & UGC 5456 & u3n8020dr + er      & 500 +    600          \\
F336W & WFPC2 $U$      & NGC 4214 & u3n80101m + 2m + 3m & 260 +    900 +    900 \\
      &                & UGC 685  & u3n80301r + 2r + 3r & 300 +    900 +    900 \\
      &                & UGC 5456 & u3n80208r + 9r + ar & 300 + 1\,000 + 1\,000 \\
F555W & WFPC2 $V$      & NGC 4214 & u3n80104m + 5m + 6m & 100 +    600 +    600 \\
      &                & UGC 685  & u3n80304r + 5r + 6r & 100 +    500 +    500 \\
      &                & UGC 5456 & u3n80207r + br + cr & 100 +    500 +    500 \\
F702W & WFPC2 wide $R$ & NGC 4214 & u3n80107m + 8m      & 500 +    500          \\
      &                & UGC 685  & u3n80307r + 8r + 9r & 100 +    500 +    500 \\
      &                & UGC 5456 & u3n80201r + 2r + 3r & 100 +    500 +    500 \\
F814W & WFPC2 $I$      & NGC 4214 & u3n8010am + bm + cm & 100 +    600 +    600 \\
      &                & UGC 685  & u3n8030ar + br + cr & 100 +    600 +    600 \\
      &                & UGC 5456 & u3n80204r + 5r + 6r & 100 +    600 +    600 \\
\enddata
\end{deluxetable}

\begin{deluxetable}{lclllll}
\tablecaption{Results for the three galaxies. In each case the position and
uncertainty of the TRGB, $m_{\rm F814W,TRGB}$, is given using both F555W$-$F814W 
and F702W$-$F814W data to discriminate between red and blue stars. The value of
$m_{\rm F814W,TRGB}$ does not include the extinction and filter transformation
corrections. For NGC 4214, results are shown both
for the PC and the WF fields. The last column shows the proposed distance for
the galaxies. As described in the text, the location of the TRGB for UGC 5456 
cannot be conclusively determined from the data and the two possible locations
lead to two different distances to the galaxy.\label{distances}}
\tabletypesize{\small}
\tablewidth{0pt}
\tablehead{\colhead{Galaxy} & \colhead{Chip(s)} & 
\multicolumn{2}{c}{F555W$-$F814W} & \multicolumn{2}{c}{F702W$-$F814W} & 
\colhead{Distance} \\
 & & \colhead{$m_{\rm F814W,TRGB}$} & \colhead{$m-M$} & 
\colhead{$m_{\rm F814W,TRGB}$} & \colhead{$m-M$} &\colhead{(Mpc)} }
\startdata
NGC 4214 & PC  & $23.48\pm 0.06$  & $27.34\pm 0.13$  & $23.48\pm 0.04$  & $27.34\pm 0.13$  &                 \\
         & WF  & $23.45\pm 0.13$  & $27.31\pm 0.18$  & $23.48\pm 0.10$  & $27.34\pm 0.16$  & $2.94\pm 0.18$  \\
UGC 685  & All & $24.48\pm 0.10$  & $28.32\pm 0.14$  & $24.56\pm 0.13$  & $28.40\pm 0.14$  & $4.79\pm 0.30$  \\
UGC 5456 & All & $24.20\pm 0.14$? & $27.87\pm 0.28$? & $24.24\pm 0.10$? & $27.91\pm 0.27$? & $3.82\pm 0.48$? \\
         &     &                  &                  & $\gtrsim 25.07$? & $\gtrsim 28.74$? & $\gtrsim 5.60$? \\
\enddata
\end{deluxetable}

\begin{deluxetable}{lcccc}
\tablecaption{Measured values, literature data, and adopted values for the
extinction. Our measured values are obtained from WFPC2 $UVRI$ photometry of 
early-type stars.\label{extinctions}}
\tablewidth{0pt}
\tablehead{\colhead{Galaxy} & \multicolumn{3}{c}{$E(B-V)$} & 
\colhead{Adopted $A_I$} \\
 & \colhead{measured} & \colhead{SFD98$^a$} & \colhead{other} & }
\startdata
NGC 4214 & 0.09 & 0.02 & $0.0-0.6^b$ & $0.11\pm 0.07$ \\
UGC 685  & 0.08 & 0.06 & $0.23^c$    & $0.13\pm 0.02$ \\
UGC 5456 & 0.27 & 0.04 &             & $0.30\pm 0.22$ \\
\enddata
\par
\begin{tabular}{l}
$^a$ \citet{Schletal98} \\
$^b$ \citet{Maizetal98} \\
$^c$ \citet{Jansetal00} \\
\end{tabular}
\end{deluxetable}

\begin{figure}
\centerline{\includegraphics*[width=\linewidth]{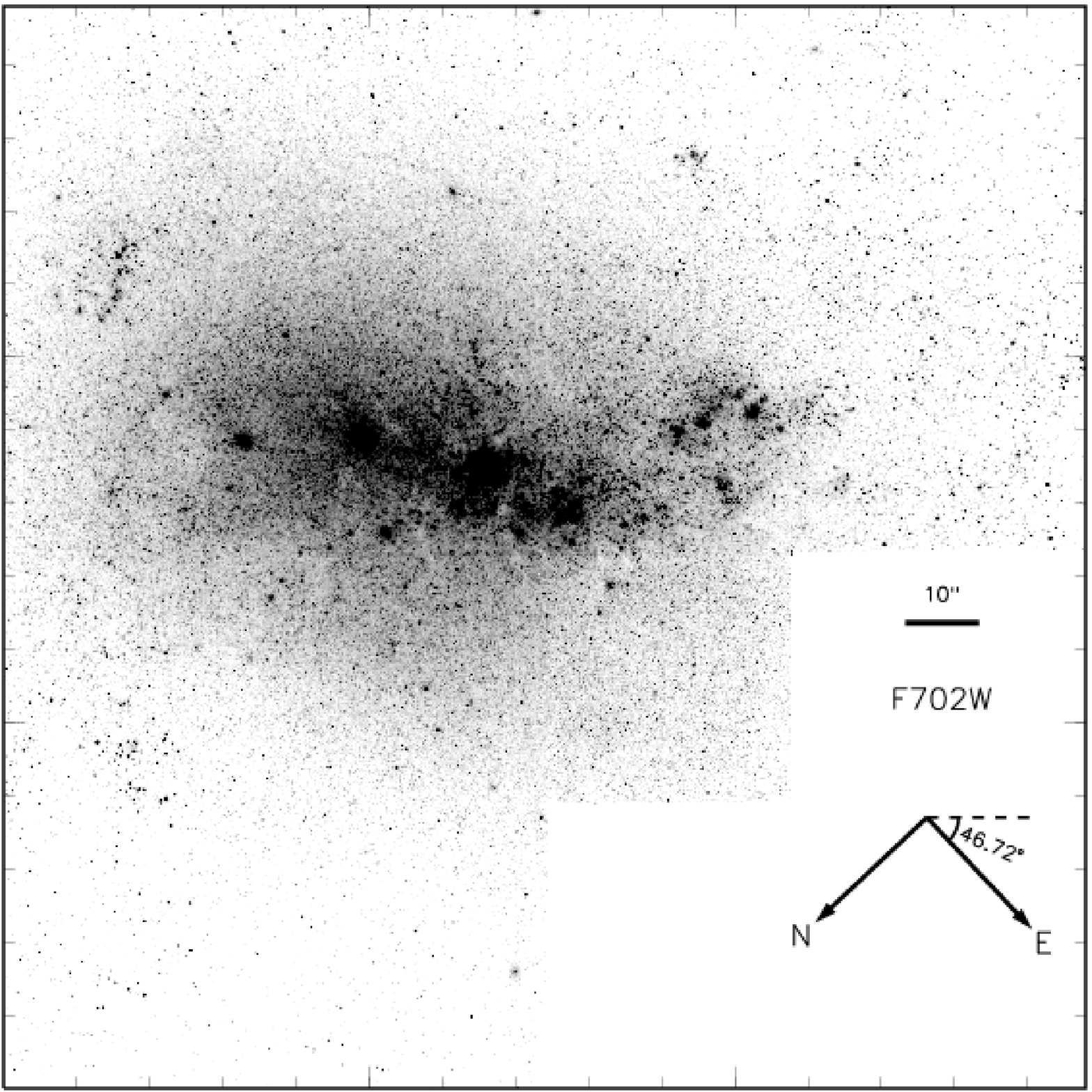}}
\caption{\ha\ subtracted WFPC2 wide $R$ image of NGC~4214. Tick marks are shown 
for every hundredth pixel.}
\label{wfpc2ngc4214}
\end{figure}

\begin{figure}
\centerline{\includegraphics*[width=\linewidth]{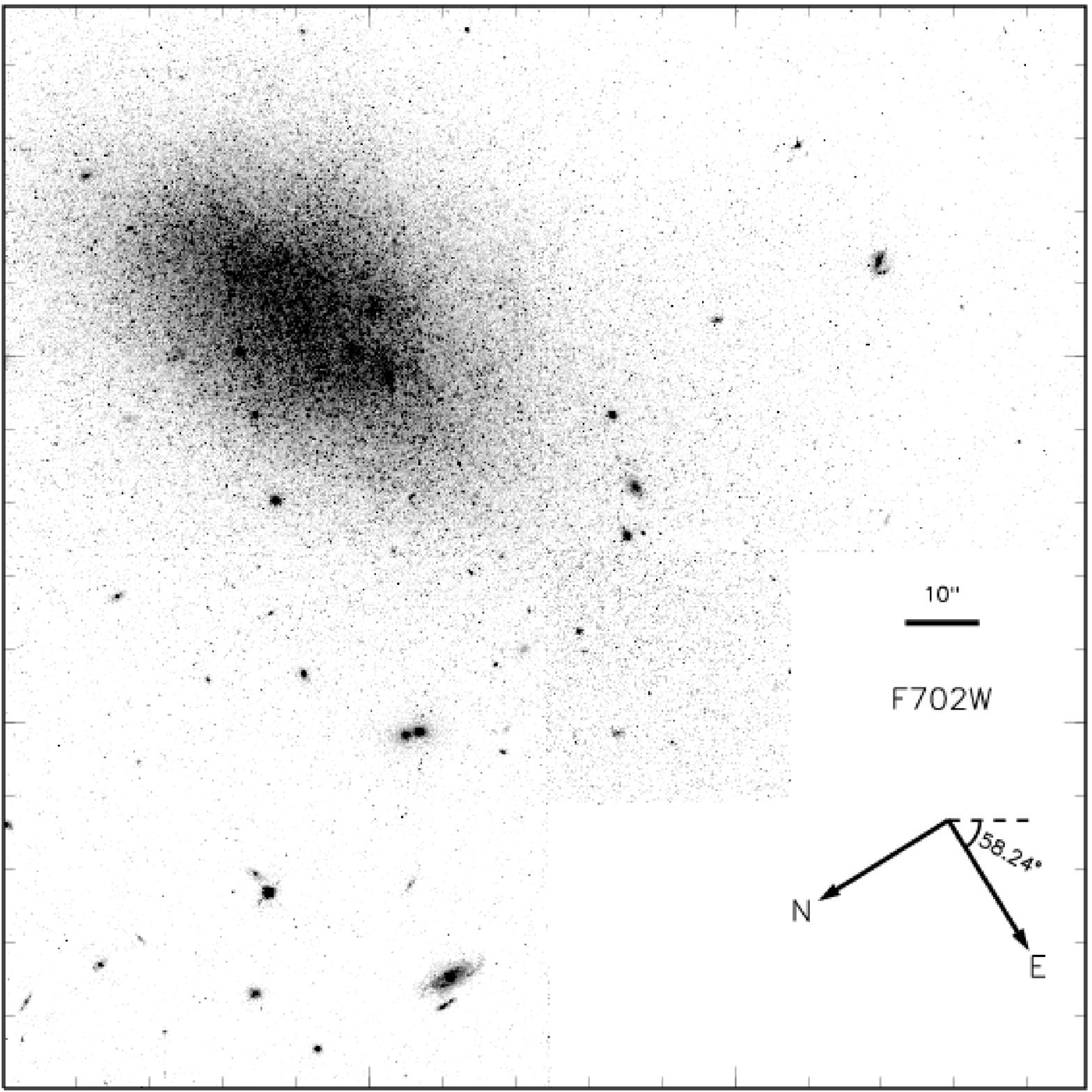}}
\caption{\ha\ subtracted WFPC2 wide $R$ image of UGC~685. Tick marks are shown 
for every hundredth pixel.}
\label{wfpc2ugc685}
\end{figure}

\begin{figure}
\centerline{\includegraphics*[width=\linewidth]{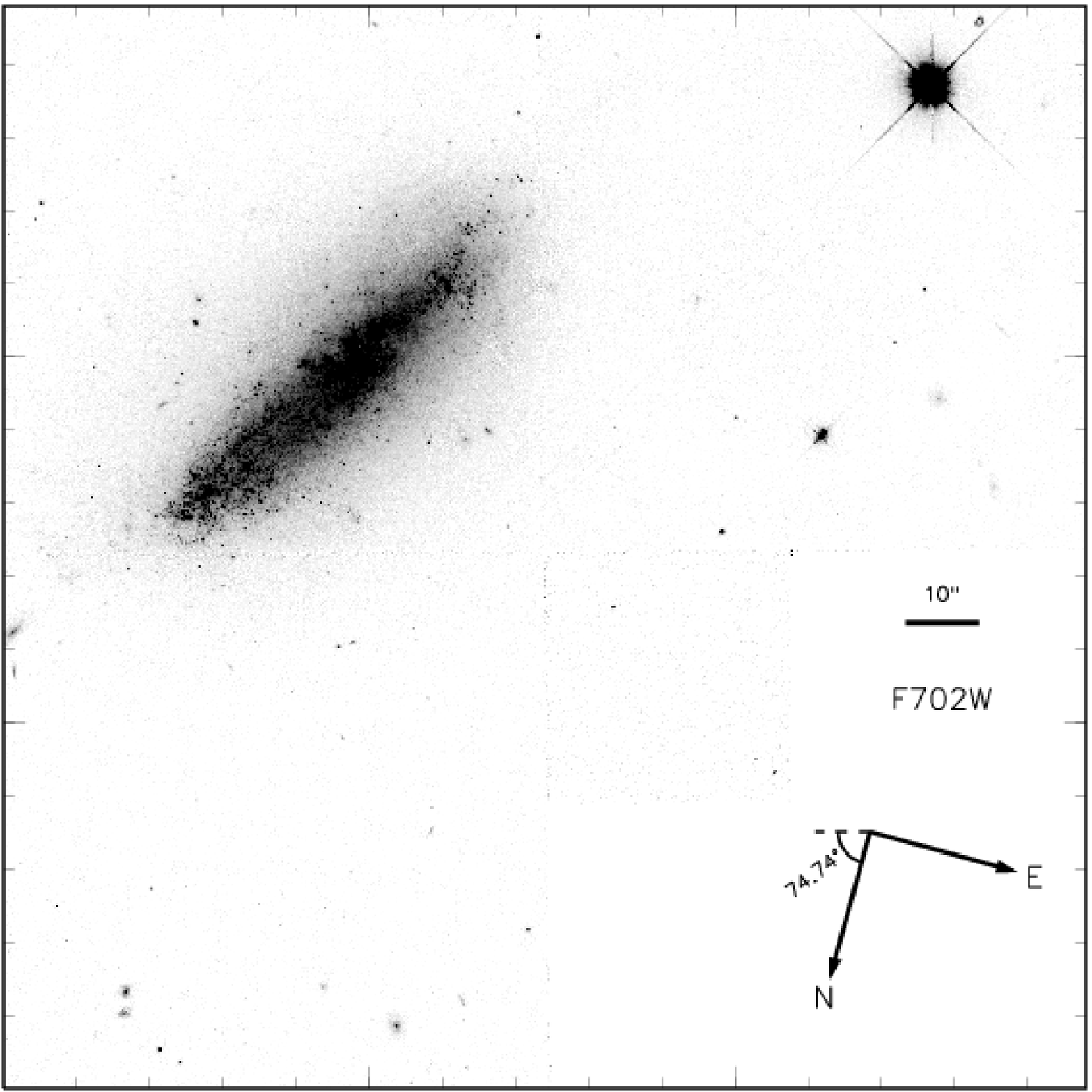}}
\caption{\ha\ subtracted WFPC2 wide $R$ image of UGC~5456. Tick marks are shown 
for every hundredth pixel.}
\label{wfpc2ugc5456}
\end{figure}

\end{document}